\documentstyle[mprocl]{article}

\def\Journal#1#2#3#4{{#1} {\bf #2}, #3 (#4)}

\def\PRL{\em Phys. Rev. Lett.}
\def\PRD{{\em Phys. Rev.} D}

\def\ep{\epsilon}
\def\ra{\rightarrow}

\def\al{\alpha}

\def\be{\begin{equation}}
\def\ee{\end{equation}}
\def\bea{\begin{eqnarray}}
\def\eea{\end{eqnarray}}

\begin{document}

\title{CAN A SELF-GRAVITATING THIN COSMIC STRING \\
OBEY THE NAMBU-GOTO DYNAMICS ?}

\author{ B. BOISSEAU, C. CHARMOUSIS, B. LINET}

\address{Laboratoire de Math\'ematiques et Physique Th\'eorique \\
CNRS/UPRES-A 6083, Universit\'e Fran\c{c}ois Rabelais \\
Parc de Grandmont 37200 Tours, France}


\maketitle\abstracts{We assume that a self-gravitating string is locally
described by a thin tube of matter represented by a 
``smoothed conical metric''. If we impose a specific constraint on the model
of string then its central line obeys the Nambu-Goto dynamics in the
limit where the radius of the tube tends to zero. If no constraint is
added then the world sheet of the central line is totally geodesic.}

It is admitted that an infinitely thin test cosmic string in a given background
spacetime obeys the Nambu-Goto dynamics as it is the case in the Minkowski
spacetime. But
there is no proof that a self-gravitating cosmic string has the same 
behaviour. Since the exterior metric 
generated by a straight cosmic string is a conical metric \cite{vil}, 
some previous works \cite{vic}$^{\!-\,}$\cite{cla}
have considered a world sheet whose the points are 
conical singularities in order to describe a cosmic string of arbitrary shape.
They have then found that this world sheet is totally geodesic which is
a very particular case of the history of a Nambu-Goto string.

The aim of this communication is to examine again the dynamics of a 
self-gravitating string within a method in which the string is represented by
a tube of matter having a thichness. We shall give a short account on the 
equations of motion of the central line of the tube of matter in the limit
where the thichness tends to zero. More complete explanations can be found in 
our paper \cite{boi}.

As a preliminary, we portray a straight cosmic string as a cylinder of matter 
in general relativity \cite{mar}$^{\!-\,}$\cite{lin}. In the coordinate 
system $(t,z,l, \varphi )$ with $l \geq 0$ and $0\leq \varphi < 2\pi$,
the interior metric is
\be
\label{1}
ds^{2}_{{\rm int}}=-dt^{2}+dz^{2}+dl^{2}+f^{2}(l)d\varphi^{2} 
\quad 0\leq l <l_{0} 
\ee
with $f\sim l+a_{3}l^{3}+\cdots$ as $l\ra 0$ to ensure a regular behaviour of
metric~(\ref{1}) at $l=0$; the exterior metric is
\be
\label{2}
ds^{2}_{{\rm ext}}=-dt^{2}+dz^{2}+dl^{2}+\sin^{2}\al (l-\overline{l}_{0})
d\varphi^{2}\quad l>l_{0}
\ee
which is the so-called conical metric. We can visualize the two-surface
$t=$ const. and $z=$ const. as a cone of half angle $\al$ whose the top
is cut out at $l_{0}-\overline{l}_{0}$ of its vertex and replaced by a cap
joining tangently the cone at $l=l_{0}$. This solution is called the
``smoothed conical metric''.

A well known solution obtained for the cylinder of
matter with a constant energy density is $f(l)=\ep \sin (l/\ep )$ which 
represents a spherical cap of radius $\ep$. This suggest to impose for the
interior metric~(\ref{1}) the form
\be
\label{3}
f(l)=\ep h\left( \frac{l}{\ep}\right) \, .
\ee
Then, the matching conditions at $l=l_{0}$ are written
\be
\label{4}
h\left( \frac{l_{0}}{\ep}\right) =\cos \al  \quad {\rm and}\quad
h'\left( \frac{l_{0}}{\ep}\right) =\sin \al
\ee
where the prime denotes the derivative.
Hence, for a given function $h$, $l_{0}/\ep$ depends only on the
angle $\al$. We can choose $\ep$ as little as we want with $l_{0}/\ep$ 
constant.

Finally, we shall also require the continuity of the second derivative of
the components of the metric at $l=l_{0}$ i.e. $h''(l_{0}/\ep )=0$.
As a consequence the Ricci tensor is continuous at $l=l_{0}$.

In a model of a self-gravitating string of arbitrary shape, we substitute
the general form $g_{AB}(\tau^{C},l,\varphi )d\tau^{A}d\tau^{B}$ for
$-dt^{2}+dz^{2}$ in the interior metric~(\ref{1}) and 
the exterior metric (\ref{2}) where the coordinates $\tau^{C}$
parametrize the world sheet of the central line $l=0$ of the tube of matter. We
suppose also that the string is like a straight one in a small
neigbhourhood and consequently it is locally characterized by a
``smooth conical metric'' given by the second part of the interior 
metric~(\ref{1}) and of the exterior metric~(\ref{2}). 
We also omit the cross terms.

We introduce the well defined coordinates $\rho^{c}$ by 
$\rho^{1}=l\cos \varphi$ and $\rho^{2}=l\sin \varphi$ which turn out to be
geodesic coordinates, for example \cite{boi2}. Then, the metric is now written
\be
\label{5}
ds^{2}=g_{AB}(\tau^{C},\rho^{c})d\tau^{A}d\tau^{B}+g_{ab}(\rho^{c})
d\rho^{a}d\rho^{b}
\ee
with $g_{ab}$ given by
\be
\label{5a}
g_{ab}(\rho^{c})=\delta_{ab}+q\left( \frac{l}{\ep}\right) \varepsilon_{ca}
\varepsilon_{db}\frac{\rho^{c}\rho^{d}}{\ep^{2}} \quad  
{\rm with} \quad q(x)=\frac{1}{x^{4}}h^{2}(x)-\frac{1}{x^{2}}
\ee
where $\varepsilon_{ab}$ is the totally antisymmetric Levi-Civita symbol.
The extrinsic curvature $K_{dAB}$ of the world sheet appears in the
expansion of the metric components
\be
\label{5b}
g_{AB}(\tau^{C},\rho^{c})=\gamma_{AB}(\tau^{C})+2K_{dAB}(\tau^{C})\rho^{d}+
O(\mid \rho^{c}\! \mid^{2}) \, .
\ee

Since the Ricci tensor is continuous at $l=l_{0}$, the interior Einstein
equations coincide with the vacuum ones at $l=l_{0}$. The method consists in
an expansion of the Ricci tensor in powers of $1/\ep$ at $l=l_{0}$ 
keeping $l_{0}/\ep$ constant. The terms in $1/\ep^{2}$ 
of $R_{\mu \nu}\mid_{l=l_{0}}$ cancel identically
since they describe the straight case. Finally, we have
\be
\label{6}
R_{\mu \nu}\mid_{l=l_{0}}\equiv R_{\mu \nu}\left( \frac{1}{\ep}\right) +
R_{\mu \nu}\left( \frac{1}{\ep^{0}}\right)=0 \, .
\ee
The leading term in $1/\ep$ in Eq.~(\ref{6}) must vanish in the limit
$\ep \ra 0$. We get
\be
\label{7}
\lim_{\ep \ra 0}\ep R_{ab}\left( \frac{1}{\ep}\right) =F_{ab}^{d}
\left( \frac{\rho_{0}^{c}}{\ep}\right) K_{d}=0 \, ,
\ee
\be
\label{8}
\lim_{\ep \ra 0}\ep R_{AB}\left( \frac{1}{\ep}\right) =F^{d}\left( 
\frac{\rho_{0}^{c}}{\ep}\right) K_{dAB}=0
\ee
where $K_{dAB}$ and $K_{d}$ are respectively the extrinsic curvature
and the mean curvature of the world sheet defined by $\rho^{c}=0$.

The functions $F_{ab}^{d}$ and $F^{d}$ depend only on the polar angle
$\varphi$ and Eq.~(\ref{7}) and Eq.~(\ref{8}) must be verified for all
$\varphi$. Therefore the generic solution is
\be
\label{9}
K_{dAB}=0
\ee
and the world sheet is totally geodesic. However another possibility exists
if $F^{d}=0$, i.e. if
\be
\label{10}
h\left( \frac{l_{0}}{\ep}\right) h'\left( \frac{l_{0}}{\ep}\right) -
\frac{l_{0}}{\ep}=0
\ee
which is an algebraic relation to be added to the boundary conditions taken
at constant $l_{0}/\ep$. We point out that an infinity of particular
solutions $h$ can be found with the aid of odd polynomials of
order greater than seven. So, Eq.~(\ref{8}) is
irrelevant but from Eq.~(\ref{7}) we obtain
\be
\label{11}
K_{d}=0
\ee
and the world sheet is a minimal surface.

In conclusion, with a reasonably general form (\ref{5}) of the metric
describing a self-gravitating string having an arbitrary shape, we have
obtained results about the equations of motion of the central line
$\rho^{c}=0$ of the tube of matter.
In the generic case, the world sheet of the central line is 
totally geodesic in the limit where the thickness of the string tends to 
zero. On the contrary, if we add the algebraic constraint (\ref{10}) then this
world sheet obeys the Nambu-Goto dynamics.

\end{document}